\documentstyle[pra,epsf,aps,multicol,cite]{revtex}

\draft
\renewcommand{\narrowtext}{\begin{multicols}{2}
\global\columnwidth20.5pc}
\renewcommand{\widetext}{\end{multicols} \global\columnwidth42.5pc}
\def\buildrel#1\under#2{\mathrel{\mathop{\kern0pt #2}\limits_{#1}}}
\def\Res#1{{\buildrel {\scriptstyle #1} \under {\textstyle \rm Res}}\,}
\begin{document}

\bibliographystyle{myprsty}

\title{Resummation of the Divergent Perturbation Series\\
for a Hydrogen Atom in an Electric Field}

\author{Ulrich~D.~Jentschura\cite{InternetULJ}}

\address{Laboratoire Kastler--Brossel, Unit\'{e} de Recherche
associ\'{e}e au CNRS N$^{\rm o}$ C8552, Case 74,\\
Universit\'{e} Pierre et Marie Curie, F-75252 Paris Cedex 05, France,\\
and National Institute of Standards and Technology,
Mail Stop 8401, Gaithersburg, Maryland 20899-8401\\
and Institut f\"ur Theoretische Physik, 
Technische Universit\"{a}t Dresden, 01062 Dresden, Germany}

\maketitle

\begin{abstract}
We consider the resummation of the perturbation series
describing the energy displacement of a hydrogenic
bound state in an electric field (known as the Stark effect or the 
LoSurdo--Stark effect), which constitutes a divergent 
formal power series in the electric field strength. 
The perturbation series
exhibits a rich singularity structure in the Borel plane.
Resummation methods are presented which 
appear to lead to consistent results even 
in problematic cases where isolated singularities
or branch cuts
are present on the positive and negative real axis in the Borel plane.
Two resummation prescriptions
are compared: (i) a variant of the Borel--Pad\'{e}
resummation method, with an additional
improvement due to utilization of the leading renormalon
poles (for a comprehensive discussion of
renormalons see [M. Beneke, Phys. Rep.~{\bf 317}, 1 (1999)]), 
and (ii) a contour-improved combination of
the Borel method with an analytic continuation by conformal 
mapping, and Pad\'{e} approximations in the conformal variable.
The singularity structure in the case of the LoSurdo--Stark effect in the 
complex Borel plane is shown to be similar to 
(divergent) perturbative expansions in quantum chromodynamics.
\end{abstract}

\noindent

\narrowtext

%\tableofcontents

\typeout{Section 1}
%
% INTRODUCTION
%
\section{INTRODUCTION}
\label{intro}

Consider the energy shift of the ground state of atomic hydrogen 
in an electric field of field strength $F$ which we 
assume to lie along the $z$-axis: the energy
displacement can be expressed
in perturbation theory as a formal power series.
The first nonvanishing
perturbation [in atomic units,
see also Eq.~(\ref{Ser000}) below] is the second-order effect 
\[
F^2 \, \sum_{m \neq \mathrm{1S}} 
\frac{\langle \phi_{\mathrm{1S}} | z | \phi_m \rangle \, 
\langle \phi_m | z | \phi_{\mathrm{1S}} \rangle}{E_{\mathrm{1S}} - E_m}\,,
\]
where the sum over $m$ runs over the entire spectrum,
including the continuum, but excluding the 1S ground state,
and $E_m$ is the nonrelativistic (Schr\"{o}dinger) energy of the $m$th state.
A well known, but perhaps surprising result says that 
the coefficients of the terms of order $F^4$, $F^6$, $F^8,\dots$
grow so rapidly that the series in $F$ ultimately diverges,
irrespective of how small the field strength is.
The convergence radius of the factorially divergent perturbation
series is zero. 
The resummation of the divergent
series is problematic in the considered case,
because the Borel transform, from which the 
physically correct, finite
result is obtained by evaluating the Laplace--Borel
integral -- see Eq.~(\ref{ContourIntegral}) in Sec.~\ref{sec_bp} below --, 
exhibits a rich singularity structure in the complex plane.

The purpose of this paper is to present numerical evidence 
that divergent perturbation series whose Borel transform
exhibits a rich singularity
structure in the complex plane, can be resummed to 
the complete, physically relevant result.
The resummation methods use as input data only a finite
number of perturbative coefficients.
Problematic singularities on the positive real
axis in the Borel plane are treated by appropriate 
integration prescriptions.
In particular, it is shown that the Borel transform of the
divergent perturbation series for the LoSurdo--Stark 
effect involves two cuts in the Borel plane, generated
essentially by the divergent alternating and nonalternating
subcomponents of the perturbation series.
This singularity structure
is also expected of the (divergent)
perturbation series in quantum field theory, notably quantum
chromodynamics (in this case, the alternating and nonalternating
factorially divergent components correspond in their mathematical 
structure to the ultraviolet and infrared
renormalons). 

We present results which suggest that the integration
contours and resummation techniques discussed here
may be of relevance, at least in part,
to theories with degenerate minima.
As a byproduct of these investigations, numerical 
pseudo-eigenvalues are obtained
for the LoSurdo--Stark effect; selected field strengths
and atomic states are considered.

The LoSurdo--Stark effect and its associated divergent 
perturbative expansion including the
the nonperturbative, nonanalytic imaginary contributions have attracted
considerable attention, both theoretically and 
experimentally~\cite{Si1978,BeGrHaSi1979,SiKo1979,Op1928,Al1969,%,
HeInBr1974,GuNi1975,FrBr1975,DaKo1976,YaTaSi1977,%,
DaKo1978,GrGr1978b,HeSi1978,%,
BeGrHaSi1979err,DaKo1979,SiAdCiOt1979,%,
AvEtAl1979,AdEtAl1980,BeGr1980,LKBa1980a,LKBa1980b,SiHaGr1981,%,
MaChRe1983,FaRe1983,%,
FrGrSi1985,GlNgYaNa1985,Ko1987,Ko1989,AlSi1989,%,
AlDaSi1991,Fe1992,NiTh1992,AlSi1994,ZaCiSk2000}. Experiments have
been performed in field strengths up to a couple of
MV/cm~\cite{Ko1978,BeEtAl1984,GlNa1985,RoWe1986}.
One might be tempted to say that the autoionization decay width
could be interpreted as a paradigmatic example for a nonperturbative
effect which exhibits fundamental limitations to the validity of
perturbation theory (unless the perturbative 
expansion is combined with resummation methods). 
We briefly summarize here:
The Rayleigh--Schr\"{o}dinger perturbation series for the
LoSurdo--Stark effect~\cite{Su1913,St1913} can be formulated to
arbitrarily high order~\cite{Si1978}. The perturbative coefficients
grow factorially in absolute magnitude~\cite{BeGrHaSi1979}, and the
radius of convergence of the perturbation series about the origin is
zero.  The perturbation series is a divergent, asymptotic expansion in
the electric field strength $F$, i.e.~about zero electric field.  This
means that the perturbative terms at small coupling first decrease in
absolute magnitude up to some minimal term.  After passing through the
minimal term, the perturbative terms increase again in magnitude, and
the series ultimately diverges.

By the use of a {\em resummation} method, it is possible to assign a
finite value to an otherwise divergent series, and various
applications of resummation methods in mathematics and physics have
been given, e.g., in~\cite{BeOr1978,We1989,BaGr1996,DuHa1999,JeBeWeSo2000}.  
When a divergent series is resummed, 
the superficial precision limit set by the 
minimal term can be overcome,
and more accurate results can be obtained as compared to the 
optimal truncation of the perturbation series
(see also the numerical results in the Tables of~\cite{Je2000prd}).
The divergent perturbation series of the LoSurdo--Stark effect has
both alternating and nonalternating components (as explained
in Sec.~\ref{sec_ps} below). The resummation of
nonalternating series or of series which have a leading or subleading
divergent nonalternating component, corresponds to a resummation ``on
the cut'' in the complex plane~\cite{BeOr1978,We1989}.
 
Rather mathematically motivated investigations regarding the Borel summability
of the divergent perturbation series for the LoSurdo--Stark effect were
performed in~\cite{GrGr1978b} and~\cite{CaGrMa1993}, and it 
was established that
the perturbation series of the LoSurdo-Stark effect
is Borel summable in the distributional
sense (for the definition of ``distributional Borel summability''
we refer to~\cite{CaGrMa1986}).  
Here, to supplement the mathematically
motivated investigations, we consider the calculation
of transforms of the divergent series, 
which use as input data only a finite number of perturbative coefficients
and exhibit apparent convergence to the 
complete, physically relevant result.

In the remarkable investigation~\cite{FrGrSi1985},
whose significance may not have been sufficiently
noticed in the field of large-order perturbation theory,
it was not only shown that it is
possible to perform the required analytic continuation of the
Borel transform beyond its radius of convergence by employing Pad\'{e}
approximants, but that it is
also possible to reconstruct the full physical result,
including the imaginary contribution which corresponds to the
autoionization decay width, by integration 
in the complex plane. 

In Sec.~\ref{sec_ps}, we discuss the 
singularity structure of the Borel transform in the complex plane. 
The structure of a doubly-cut plane has been postulated for
quantum chromodynamic perturbation series~\cite{CaFi1999,CaFi2000},
and this structure has been exploited to devise resummation
prescriptions based on 
conformal mappings~\cite{CaFi1999,CaFi2000,SeZJ1979,LGZJ1983,%,
GuKoSu1995,GuKoSu1996,JeSo2000hep}.
Here, we present results which suggest that the 
convergence of the transforms obtained by conformal mapping 
can be improved if Pad\'{e} approximants in the conformal variable
are used (see also~\cite{JeSo2000hep}). We also discuss improvements
of the ``pure'' Borel--Pad\'{e} method (these additional improvements
take advantage of leading renormalon poles).
Also, in comparison to the investigation~\cite{FrGrSi1985},
we use here a slightly modified, but equivalent integration contour
for the evaluation of the generalized Borel integral
(see~\cite{Je2000prd} and Sec.~\ref{sec_bp} below).
Our version of the integration contour
exhibits the additional terms which have to be added to the otherwise
recommended principal-value
prescription~\cite{Pi1999,CaFi1999,CaFi2000}. 

As stressed above, it has been
another main motivation for the current investigation
to establish the singularity structure of the Borel transform
in the complex plane, and to demonstrate the analogy of 
the singularity structure of the 
perturbation series for the LoSurdo--Stark effect
to quantum chromodynamic perturbation series.     
We also consider a divergent perturbation
series generated by a model problem for 
theories with degenerate minima.
In the particular model case discussed here,
a perturbation series with real coefficients is summed 
to a {\em real} result -- in contrast
to the LoSurdo--Stark effect, there is 
no imaginary part involved in this case.
One of the three alternative integration contours introduced 
in~\cite{Je2000prd} has to be employed.

This paper is organized as follows: In Sec.~\ref{sec_ps}, we give
a brief outline of the perturbative expansion
for the LoSurdo--Stark effect. In Secs.~\ref{sec_bp} and~\ref{sec_cm},
we describe the resummation methods which are used to obtain
the numerical results presented in Sec.~\ref{sec_nc}.
In Sec.~\ref{degen}, we consider theories with degenerate minima.
We conclude with a summary of the results in Sec.~\ref{sec_co}.
Finally, the connection of the current investigation to
quantum field theoretic perturbation series and to double-well
oscillators are discussed in the
Appendixes~\ref{appa} and~\ref{appb}.

\typeout{Section 2}
%
% PERTURBATION SERIES FOR THE LoSURDO--STARK EFFECT
%
\section{PERTURBATION SERIES\\ FOR THE L\lowercase{o}SURDO--STARK EFFECT}
\label{sec_ps}

In the presence of an electric field, the ${\rm SO}(4)$ symmetry of
the hydrogen atom is broken, and parabolic quantum numbers $n_1$,
$n_2$ and $m$ are used for the classification of the atomic
states~\cite{LaLi1979}. For the Stark effect, the perturbative
expansion of the energy eigenvalue $E(n_1,n_2,m,F)$ reads [see
Eq.~(59) of~\cite{Si1978}],
\begin{equation}
\label{PertSer}
E(n_1,n_2,m,F) \sim \sum_{N=0}^{\infty}
E^{(N)}_{n_1 n_2 m} \, F^N\,,
\end{equation}
where $F$ is the electric field strength.  For $N \to \infty$, the leading
large-order factorial asymptotics of the perturbative coefficients have been
derived in~\cite{SiAdCiOt1979} as
\begin{equation}
\label{LargeN}
E^{(N)}_{n_1 n_2 m} \sim A^{(N)}_{n_1 n_2 m} +
(-1)^N \, A^{(N)}_{n_2 n_1 m}\,, \quad N \to \infty\,,
\end{equation}
where $A^{(N)}_{n_i n_j m}$ is given as an asymptotic series,
\begin{eqnarray}
\label{DefAN}
\lefteqn{A^{(N)}_{n_i n_j m} \sim K(n_i,n_j,m,N)}
\nonumber\\[1ex]
& & \quad \times
\sum_{k=0}^{\infty} a^{n_i n_j m}_k \, (2\,n_j + m + N - k)!\,.
\end{eqnarray}
The quantities $a^{n_i n_j m}_k$ are constants.
The $K$-coefficients in Eq.~(\ref{DefAN}) are given by
\begin{eqnarray}
\label{Prefactor}
\lefteqn{K(n_i,n_j,m,N) = - \left[ 2\pi n^3 n_j! \, (n_j + m)! \right]^{-1}}
\nonumber\\[2ex]
& & \quad \times \exp\left\{ 3 \, (n_i - n_j) \right\} \,
6^{2 \, n_j + m + 1} \, (3 n^3/2)^N \,.
\end{eqnarray}
Here, the principal quantum number $n$ as a function of the parabolic
quantum numbers $n_1$, $n_2$ and $m$ is given by [see Eq.~(65)
in~\cite{Si1978}]
\begin{equation}
\label{PrincQuant}
n = n_1 + n_2 + |m| + 1\,.
\end{equation}
According to Eq.~(\ref{LargeN}), the perturbative coefficients
$E^{(N)}_{n_1 n_2 m}$, for large order $N \to \infty$ of perturbation
theory, can be written as a sum of a nonalternating factorially
divergent series [first term in Eq.~(\ref{LargeN})] and of an
alternating factorially divergent series [second term in
Eq.~(\ref{LargeN})].  Because the $a^{n_i n_j m}_k$ in
Eq.~(\ref{DefAN}) are multiplied by the factorial $(2\,n_i + m + N -
k)!$, we infer that for large perturbation theory order $N$, the term
related to the $a^{n_i n_j m}_0$ coefficient ($k = 0$) dominates.
Terms with $k \geq 1$ are suppressed in relation to the leading term
by a relative factor of $1/N^k$ according to the asymptotics
\begin{equation}
\label{AsyN}
\frac{(2\,n_j + m + N - k)!}{(2\,n_j + m + N)!} \sim
\frac{1}{N^k} \, \left[ 1 + {\mathcal O}\left(\frac{1}{N}\right)\right]\,
\end{equation}
for $N \to \infty$. The leading ($k=0$)--coefficient has been evaluated
in~\cite{BeGrHaSi1979} as
\begin{equation}
\label{ninjm0}
a^{n_i n_j m}_0 = 1\,.
\end{equation}
According to Eqs.~(\ref{LargeN}), (\ref{DefAN}) and (\ref{ninjm0}),
for states with $n_1 < n_2$, the nonalternating component of the
perturbation series dominates in large order of perturbation theory,
whereas for states with $n_1 > n_2$, the alternating component is
dominant as $N \to \infty$. For states with $n_1 = n_2$, the odd-$N$
perturbative coefficients vanish~\cite{SiAdCiOt1979}, and the even-$N$
coefficients necessarily have the same sign in large order [see
Eq.~(\ref{LargeN})]. According to Eq.~(\ref{LargeN}), there are
subleading divergent nonalternating contributions for states with $n_1
> n_2$, and there exist subleading divergent alternating contributions
for states with $n_1 < n_2$. This complicates the resummation of the
perturbation series.

%
% figure 1
%
\begin{figure}
\begin{center}
\begin{minipage}{8cm}
\centerline{\mbox{\epsfysize=5.5cm\epsffile{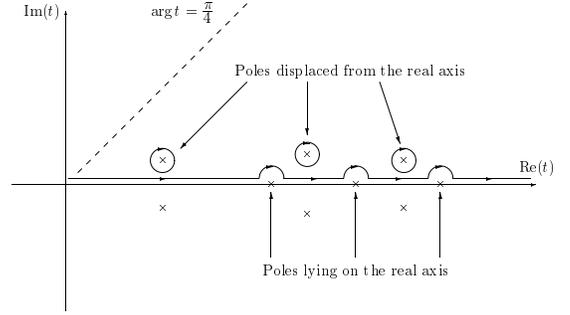}}}
\caption{\label{figg1} Integration contour
$C_{+1}$ for the evaluation of the
generalized Borel integral defined
in Eq.~\protect{(\ref{ContourIntegral})}.
Poles displaced from the real axis are evaluated
as full poles, whereas those poles which lie on the
real axis are treated as half poles.}
\end{minipage}
\end{center}
\end{figure}

\typeout{Section 3}
%
% BOREL--PAD\'{E} RESUMMATION
%
\section{BOREL--PAD\'{E} RESUMMATION}
\label{sec_bp}

The resummation of the perturbation series~(\ref{PertSer}) 
by a combination of the Borel transformation 
and Pad\'{e} approximants proceeds as
follows. First we define the parameter
\begin{equation}
 \label{deflambda}
\lambda = 2 \, {\rm max}(n_1,n_2) + m + 1\,.
\end{equation}
The large-order growth of the perturbative coefficients [see
Eqs.~(\ref{LargeN}) and (\ref{DefAN})] suggests the definition of the
(generalized) Borel transform [see Eq.~(4) in~\cite{JeWeSo2000}]
\begin{eqnarray}
\label{BorelTrans}
E_{\rm B}(z) & \equiv & E_{\rm B}(n_1,n_2,m,z) \nonumber\\[2ex]
& = & {\mathcal B}^{(1,\lambda)}\left[E(n_1,n_2,m); \, z\right] 
\nonumber\\[1ex]
& = & \sum_{N=0}^{\infty}
\frac{E^{(N)}_{n_1 n_2 m}}{\Gamma(N + \lambda)} \, z^N\,,
\end{eqnarray}
where we consider the argument $z$ of $E_{\rm B}(z)$ as a complex
variable and $\lambda$ is defined in Eq.~(\ref{deflambda}). 
The additive constant (in this case $\lambda$)
in the argument of the Gamma function
is chosen in accordance with the notion of an ``asymptotically
improved'' resummation (see also~\cite{JeWeSo2000}).
It is observed that the additive constant $\lambda$
can be shifted by a small integer without 
affecting the convergence of the Borel resummed series.
Because the perturbative coefficients $E^{(N)}_{n_1 n_2 m}$
diverge factorially in absolute magnitude, the 
Borel transform $E_{\rm B}(z)$ has a finite radius of 
convergence about the origin.  The
evaluation of the (generalized)
Laplace--Borel integral~[see Eq.~(\ref{ContourIntegral})
below] therefore requires an analytic continuation of $E_{\rm B}(z)$
beyond the radius of convergence.  The ``classical'' Borel integral is
performed in the $z$-range $z \in (0,\infty)$, i.e.~along the positive
real axis [see e.g.~Eqs.~(8.2.3) and~(8.2.4) of~\cite{BeOr1978}].  It
has been suggested in~\cite{Pi1999} that the analytic continuation
of~(\ref{BorelTrans}) into regions where $F$ retains a nonvanishing,
albeit infinitesimal, imaginary part can be achieved by evaluating
Pad\'{e} approximants.  Using the first $M+1$ terms in the power
expansion of the Borel transform $E_{\rm B}(z)$, we construct the
Pad\'{e} approximant~(we follow the notation of~\cite{BaGr1996})
\begin{equation}
\label{pade}
{\mathcal P}_M(z) = \bigg[ [\mkern - 2.5 mu [M/2] \mkern - 2.5 mu ] \bigg/
[\mkern - 2.5 mu [(M+1)/2] \mkern - 2.5 mu ]
\bigg]_{E_{\rm B}}\!\!\!\left(z\right)\,,
\end{equation}
where $[\mkern - 2.5 mu [x] \mkern - 2.5 mu ]$ denotes the largest
positive integer smaller than $x$.  We then evaluate the (modified)
Borel integral along the integration contour $C_{+1}$ shown in
Fig.~\ref{figg1} in order to construct the transform ${\mathcal T}\!E_M(F)$
where
\begin{equation}
\label{ContourIntegral}
{\mathcal T}\!E_M(F) = \int_{C_{+1}} {\rm d}t \, t^{\lambda - 1}\,
\exp(-t)\,{\mathcal P}_M(F\,t) \,.
\end{equation}
The successive evaluation of transforms ${\mathcal T}\!E_M(F)$ in increasing
transformation order $M$ is performed, and the apparent convergence of
the transforms is examined. This procedure is illustrated in Tables~I
and II of~\cite{Je2000prd}.  In the current
evaluation, a slightly modified scheme is used for selecting
the poles in the upper right quadrant of the complex plane
as compared to~\cite{Je2000prd}. 

The contour
$C_{+1}$ is supposed to 
encircle all poles at $t = z_i$ in the upper right quadrant 
of the complex plane 
with $\arg \, z_i < \pi / 4$ in the mathematically negative sense. 
That is to say,
the contribution of all poles $z_i$ with ${\rm Re}\,z_i > 0$,
${\rm Im}\,z_i > 0$ and ${\rm Re}\,z_i > {\rm Im}\,z_i$,
\[
- 2\,\pi\,{\rm i}\, \sum_i \,\Res{t=z_i} t^{\lambda - 1}\,
\exp(-t)\,{\mathcal P}_M(F \, t)\,,
\]
is added to the principal value (P.V.)
of the integral (\ref{ContourIntegral})
carried out in the range $t \in (0,\infty)$. 
Note the further restriction (${\rm Im}\,z_i < {\rm Re}\,z_i$
or equivalently $\arg \, z_i < \pi / 4$)
with regard to the selection of poles
in comparison to the previous investigation~\cite{Je2000prd}.
In practical calculations, this modification is 
observed not to affect the numerical values of the transforms
${\mathcal T}\!E_M(F)$ defined in Eq.~(\ref{ContourIntegral})
in higher transformation order $M \geq 10$ [i.e.~for large
$M$, see also Eq.~(\ref{Limit}) below], because the poles are
observed to cluster near the real axis
in higher transformation order, and so
the contribution of poles with $\pi / 4 < \arg z_i < \pi/2$
gradually vanishes.
We have,
\begin{eqnarray}
\label{ContourIntegral2}
\lefteqn{{\mathcal T}\!E_M(F) = {\rm (P.V.)}
\int_0^\infty {\rm d}t \, t^{\lambda - 1}\, \exp(-t)\,{\mathcal P}_M(F\,t)}
\nonumber\\[1ex]
& & \quad - 2\,\pi\,{\rm i}\, \sum_i \,\Res{t=z_i} t^{\lambda - 1}\,
\exp(-t)\,{\mathcal P}_M(F \, t) \,.
\end{eqnarray}
The principal-value prescription [first term in 
Eq.~(\ref{ContourIntegral2})]
for the evaluation of the Laplace--Borel integral
has been recommended in~\cite{Pi1999,Raczka1991}.
This prescription leads to a real (rather than complex) result for the
energy shift and cannot account for the width of the 
quasistationary state. 
The additional pole contributions [second term in
Eq.~(\ref{ContourIntegral2})]
are responsible for complex-valued (imaginary) corrections which lead, in
particular, to the decay width.  

By contour integration (Cauchy Theorem) and Jordan's Lemma,
one can show that the result obtained along $C_{+1}$ is 
equivalent to an integration along the straight line with
$\arg z = \pi/4$,
\begin{equation}
\label{ContourIntegral3}
{\mathcal T}\!E_M(F) = c^\lambda \, \int_0^\infty 
{\rm d}t \, t^{\lambda - 1}\, \exp(-c\,t)\,{\mathcal P}_M(F \, c\, t)
\end{equation}
where $c = \exp(i\,\pi/4)$. This contour has been used 
in~\cite{FrGrSi1985} (see also p.~815 in~\cite{ZJ1996}).
The exponential factor $\exp(-c\,t)$ and the asymptotic behavior of the
Pad\'{e} approximant ${\mathcal P}_M(F \, c\, t)$ as $t \to \infty$
together ensure that the integrand falls off sufficiently rapidly so that
the Cauchy Theorem and
Jordan's Lemma can be applied to show the equivalence of the 
representations (\ref{ContourIntegral2}) and (\ref{ContourIntegral3}).

The representation (\ref{ContourIntegral3}) illustrates the fact
that the integration in the complex plane along $C_{+1}$ analytically
continues the resummed result in those cases where the evaluation of the
standard Laplace--Borel integral is not feasible due to poles on the
real axis. The representations (\ref{ContourIntegral}) and
(\ref{ContourIntegral2}) serve to clarify the role of the additional
terms which have to be added to the result obtained by the principal-value
prescription in order to obtain the full physical result, including
the nonperturbative, nonanalytic contributions. Note that, as 
stressed in~\cite{Je2000prd}, the pole contributions in general do not
only modify the imaginary, but also the real part of the resummed
value for the perturbation series.

Formally, the limit of the sequence of the
${\mathcal T}\!E_M(F)$ as $M\to\infty$, provided it exists,
yields the nonperturbative result
inferred from the perturbative expansion~(\ref{PertSer}),
\begin{equation}
\label{Limit}
\lim_{M \to \infty} {\mathcal T}\!E_M(F) = E(F) \equiv 
E(n_1,n_2,m,F)\,.
\end{equation}
Because the contour $C_{+1}$ shown in Fig.~\ref{figg1} extends into
the complex plane, the transforms ${\mathcal T}\!E_M(F)$ acquire an
imaginary part even though the perturbative coefficients in
Eq.~(\ref{PertSer}) are real.

In the context of numerical analysis, the concept of
incredulity~\cite{Wi1966} may be used for the analysis of the
convergence of the transforms ${\mathcal T}\!E_M(F)$ of increasing order
$M$. If a certain number of subsequent transforms exhibit apparent
numerical
convergence within a specified relative accuracy, then the calculation
of transforms is stopped, and the result of the last calculated
transformation is taken as the numerical limit of the series under
investigation.  It has been observed in~\cite{Je2000prd,Pi1999} that for
a number of physically relevant perturbation series, the apparent
numerical convergence of the transforms (\ref{ContourIntegral}), with
increasing transformation order, leads to the physically correct
results. 

It is observed that the rate of convergence of the transforms 
(\ref{ContourIntegral}) can be enhanced if instead of the 
unmodified Pad\'{e} approximants (\ref{pade}) leading 
``renormalon'' poles are explicit used for the construction
of modified approximants. For the ground state,
this entails the following replacement in
Eq.~(\ref{ContourIntegral}):
\[
{\mathcal P}_M(z) \to {\mathcal P'}_M(z)\,,
\]
where
\begin{equation}
\label{replacement000}
{\mathcal P'}_M(z) = \frac{1}{1-z^2}\,
\bigg[ [\mkern - 2.5 mu [\frac{M+4}{2}] \mkern - 2.5 mu ] \bigg/
[\mkern - 2.5 mu [\frac{M-3}{2}] \mkern - 2.5 mu ]
\bigg]_{E'_{\rm B}(\zeta)}\!\!\!\left(z\right)\,,
\end{equation}
where $E'_{\rm B}(\zeta) = (1-\zeta^2)\,E_{\rm B}(\zeta)$.
For the excited state with quantum numbers $n_1 = 3$,
$n_2 = 0$ and $m = 1$, we replace
\[
{\mathcal P}_M(z) \to {\mathcal P''}_M(z)\,,
\]
where
\begin{equation}
\label{replacement301}
{\mathcal P''}_M(z) = \frac{1}{1-z}\,
\bigg[ [\mkern - 2.5 mu [\frac{M+2}{2}] \mkern - 2.5 mu ] \bigg/
[\mkern - 2.5 mu [\frac{M-1}{2}] \mkern - 2.5 mu ]
\bigg]_{E''_{\rm B}}\!\!\!\left(z\right)\,,
\end{equation}
where $E''_{\rm B}(\zeta) = (1-\zeta^2)\,E_{\rm B}(\zeta)$.
The resummation method by a combination of Borel and Pad\'{e}
techniques -- current Section -- will be
referred to as ``method I'' throughout the current paper.

\typeout{Section 4}
%
% DOUBLY--CUT BOREL PLANE AND RESUMMATION BY CONFORMAL MAPPING
%
\section{DOUBLY--CUT BOREL PLANE\\
AND RESUMMATION\\ BY CONFORMAL MAPPING}
\label{sec_cm}

According to Eqs.~(\ref{LargeN}) and (\ref{DefAN}), the perturbative
coefficient $E^{(N)}_{n_1 n_2 m}$, for large $N$, can be written as
the sum of an alternating and of a nonalternating divergent series. In
view of Eqs.~(\ref{Prefactor}) and (\ref{ninjm0}), we conclude
that the series defined in~Eq.~(\ref{BorelTrans}),
\[          
E_{\rm B}(z) = \sum_{N=0}^{\infty}    
\frac{E^{(N)}_{n_1 n_2 m}}{\Gamma(N + \lambda)} \, z^N\,,
\]
has a radius of convergence 
\begin{equation}   
\label{defS}
s = \frac{2}{3 \, n^3}
\end{equation}  
about the origin, where $n$ is the principal quantum number
[see Eq.~(\ref{PrincQuant})]. Therefore, the function
\begin{equation}      
\label{calEB}
{\mathcal E}_{\rm B}(w) = \sum_{N=0}^{\infty}
\frac{E^{(N)}_{n_1 n_2 m}\,\,s^N}{\Gamma(N + \lambda)} \, w^N\,, 
\end{equation}
has a unit radius of convergence about the origin.  It is not {\em a
  priori} obvious if the points $w = -1$ and $w = +1$ represent
isolated singularities or branch points.  The asymptotic properties
(\ref{LargeN}) and (\ref{DefAN}) together with Eq.~(\ref{AsyN})
suggest that the points $w = -1$ and $w = +1$ do not constitute poles
of finite order.  We observe that the leading factorial growth of the
perturbative coefficients in large perturbation order $N$ is divided
out in the Borel transform (\ref{calEB}), which is a sum over $N$. The
perturbative coefficient $E^{(N)}_{n_1 n_2 m}$ can be written as an
asymptotic series over $k$ [see Eq.~(\ref{DefAN})].  We interchange
the order of the summations over $N$ and $k$, we use Eq.~(\ref{AsyN})
and take advantage of the identity
\begin{equation}
\label{polylog}
\sum_{N=0}^{\infty} \frac{w^N}{N^k} = {\rm Li}_k (w)\,.
\end{equation}
The Borel transform ${\mathcal E}_{\rm B}(w)$ can then be written
as a sum over terms of the form $T_k(w)$ where for $k \to \infty$,
\begin{equation}
T_k(w) \sim C(n_i,n_j,m) \, a_k^{n_i n_j m} \, {\rm Li}_k (w)\,.
\end{equation}
The coefficient $C(n_i,n_j,m)$ is given by
\begin{eqnarray}
\lefteqn{C(n_i,n_j,m) = - \left[ 2\pi n^3 n_j! \, (n_j + m)! \right]^{-1}}
\nonumber\\[2ex]
& & \quad \times
\exp\left\{ 3 \, (n_i - n_j) \right\} \, 6^{2 \, n_j + m + 1} \,.
\end{eqnarray}
These considerations suggest that the points $w = -1$ and $w = +1$
represent essential singularities (in this case, branch points) of the
Borel transform ${\mathcal E}_{\rm B}(w)$ defined in Eq.~(\ref{calEB}).
For the analytic continuation of ${\mathcal E}_{\rm B}(w)$ by conformal
mapping, we write $w$ as
\begin{equation}      
\label{conformal1}
w = \frac{2\,y}{1 + y^2}
\end{equation}
(this conformal mapping preserves the origin of the complex plane).
Here, we refer to $w$ as the Borel variable, and we call $y$ the
conformal variable.  We then express the $M$th partial sum of
(\ref{calEB}) as
\begin{eqnarray}
\label{calEBM}
\lefteqn{{\mathcal E}^{M}_{\rm B}(w) = \sum_{N=0}^{M}
\frac{E^{(N)}_{n_1 n_2 m}\,s^N}{\Gamma(N + \lambda)} \, w^N}
\nonumber\\[1ex]
& & \quad = \sum_{N=0}^{M} C_N \, y^N + {\mathcal O}(y^{M+1})\,,
\end{eqnarray}
where the coefficients $C_N$ are uniquely determined [see, e.g.,
Eqs.~(36) and (37) of \cite{CaFi1999}].  We define the partial sum of the
Borel transform, re-expanded in terms of the conformal variable
$y$, as
\begin{equation}
\label{PartialConformal}
{\mathcal E}'^{M}_{\rm B}(y) = \sum_{N=0}^{M} C_N\,y^N\,.
\end{equation}
We then evaluate (lower-diagonal) Pad\'{e} approximants to the
function ${\mathcal E}'^{M}_{\rm B}(y)$,
\begin{equation}
\label{ConformalPade}
{\mathcal E}''^{M}_{\rm B}(y) =
\bigg[ [\mkern - 2.5 mu [M/2] \mkern - 2.5 mu ] \bigg/
[\mkern - 2.5 mu [(M+1)/2] \mkern - 2.5 mu ]
\bigg]_{{\mathcal E}'^{M}_{\rm B}}\!\!\!\left(y\right)\,.
\end{equation}
We define the following transforms,
\begin{equation}
\label{AccelTrans}
{\mathcal T}''\!E_M(F) = s^\lambda\,
\int_{C_{+1}} {\rm d}w \,w^{\lambda - 1} \,\exp\bigl(-w\bigr)\,
{\mathcal E}''^{M}_{\rm B}\bigl(y(w)\bigr)\,.
\end{equation}
At increasing $M$, the limit as $M\to\infty$, provided it exists, is
then again assumed to represent the complete, physically relevant
solution,
\begin{equation}
E(F) = \lim_{M\to\infty} {\mathcal T}''\!E_M(F)\,.
\end{equation}
We do not consider the question of the existence of this limit here
(for an outline of questions related to these issues we refer
to~\cite{CaFi2000}; potential problems at excessively strong
coupling are discussed in Sec. IIC of~\cite{LGZJ1980}).

Inverting Eq.~(\ref{conformal1}) yields [see Eq.~(\ref{AccelTrans})]
\begin{equation}
\label{conformal2}
y(w) = \frac{\sqrt{1+w}-\sqrt{1-w}}{\sqrt{1+w}+\sqrt{1-w}}\,.
\end{equation}
The conformal mapping given by Eqs.~(\ref{conformal1}) and
(\ref{conformal2}) maps the doubly cut $w$-plane with cuts running
from $w = 1$ to $w = \infty$ and $w = -1$ to $w = -\infty$ unto the
unit circle in the complex $y$-plane. The cuts themselves are mapped
to the edge of the unit circle in the $y$-plane.  

In comparison to the investigations~\cite{CaFi1999} and~\cite{CaFi2000}, 
we use here a
different conformal mapping defined in Eqs.~(\ref{conformal1})
and~(\ref{conformal2}) which reflects the different singularity
structure in the complex plane [cf.~Eq.~(27) in~\cite{CaFi1999}]. We
also mention the application of Pad\'{e} approximants for the
numerical improvement of the conformal mapping performed according to
Eq.~(\ref{ConformalPade}).  In comparison to a recent
investigation~\cite{JeSo2000hep}, where the additional
Pad\'{e}--improvement in the conformal variable is also used, we
perform here the analytic continuation by a mapping whose structure
reflects the double cuts suggested by the asymptotic properties of the
perturbative coefficients given in Eqs.~(\ref{LargeN}), (\ref{DefAN})
and (\ref{AsyN}) [cf.~Eq.~(5) in~\cite{JeSo2000hep}].

The method introduced in this Section will be referred to as ``method
II''. It is one of the motivations for the current investigation to
contrast and compare the two methods I and II.  A comparison of
different approaches to the resummation problem for series with both
alternating and nonalternating divergent components appears useful, in
part because the conformal mapping (without further Pad\'{e}
improvement) has been recommended for the resummation of quantum
chromodynamic perturbation series~\cite{CaFi1999,CaFi2000}.

We do not consider order-dependent mappings
here~\cite{SeZJ1979,LGZJ1983,GuKoSu1995,GuKoSu1996}. 
For an order-dependent mapping to be
constructed, the conformal mapping in Eq.~(\ref{conformal1}) has to be
modified, and a free parameter $\rho$ has to be introduced.  The coefficients
$C_N$ in the accordingly modified
Eq.~(\ref{PartialConformal}) then become $\rho$-dependent.
The free parameter $\rho$ is chosen so that the $\rho$-dependent
coefficient $C_M(\rho)$ 
of order $M$ vanishes.  Consequently, the choice of $\rho$
depends on the order $M$ of perturbation theory, and in this way the
mapping becomes order-dependent.  Certain complications arise because
$\rho$ cannot be chosen arbitrarily, but has to be selected, roughly
speaking, as the first zero of the $\rho$-dependent coefficient
$C_M(\rho)$ for which the absolute magnitude of the derivative
$C_M'(\rho)$ is small (this is explained in~\cite{ZJ1996}, p.~886). It
is conceivable that with a judicious choice of $\rho$, further
acceleration of the convergence can be achieved, especially when the
order-dependent mapping is combined with a Pad\'{e} approximation as it
is done here in Eq.~(\ref{ConformalPade}) with our
order-{\em independent} mapping.  In the current investigation, we
restrict the discussion to the conformal order-independent mapping
(\ref{conformal1}) which is nevertheless optimal in the sense
discussed in~\cite{CaFi1999,CaFi2000}.

\typeout{Section 5}
%
% NUMERICAL CALCULATIONS
%
\section{NUMERICAL CALCULATIONS}
\label{sec_nc}

In this section, the numerical results based on the resummation
methods introduced in Secs.~\ref{sec_bp} and~\ref{sec_cm} are
presented. Before we describe the calculation in detail, we should
note that relativistic corrections to both the real and the imaginary
part of the energy contribute at a relative order of $(Z\alpha)^2$
compared to the leading nonrelativistic effect which is treated in the
current investigation (and in the previous work on the subject, see
e.g.~\cite{SiAdCiOt1979,FrGrSi1985}). Therefore, the theoretical
uncertainty due to relativistic effects
can be estimated to be, at best, 1 part in $10^4$ (for an outline of
the relativistic and quantum electrodynamic corrections in hydrogen
see~\cite{BeLiPi1973,JoSo1985,Mo1996,JePa1996,JeSoMo1997,%,
MoPlSo1998,EiGrSh2000}).
Measurements in very high fields are difficult~\cite{Ko1978}. At the
achievable field strengths to date (less than $0.001\,{\rm a.u.}$ or
about $5\,{\rm MV}/{\rm cm}$), the accuracy of the theoretical
prediction exceeds the experimental precision, and relativistic
effects do not need to be taken into account.

The perturbative coefficients $E^{(N)}_{n_1 n_2 m}$ defined in
Eq.~(\ref{PertSer}) for the energy shift can be inferred, to
arbitrarily high order, from the Eqs.~(9), (13--15), (28--33),
(59--67) and (73) in~\cite{Si1978}.  The atomic unit system is used in
the sequel, as is customary for this type of
calculation~\cite{Si1978,HeInBr1974,DaKo1976,DaKo1978}.  The unit of
energy is $\alpha^2\,m_{\rm e}\,c^2 = 27.211\,{\rm eV}$ where $\alpha$
is the fine structure constant, and the unit of the electric field is
the field strength felt by an electron at a distance of one Bohr
radius $a_{\rm Bohr}$ to a nucleus of elementary charge, which is
$1/(4\,\pi\,\epsilon_0)\,(e/a_{\rm Bohr}^2) = 5.142\times 10^{3}\,
{\rm MV}/{\rm cm}$ (here, $\epsilon_0$ is the permittivity of the
vacuum).

We consider the resummation of the divergent perturbative
expansion~(\ref{PertSer}) for two states of atomic hydrogen. These are
the ground state ($n_1 = n_2 = m = 0$) and an excited state with
parabolic quantum numbers $n_1 = 3$, $n_2=0$, $m=1$. 
We list here the first few perturbative
coefficients for the states under investigation. For the ground state,
we have (in atomic units),
\begin{eqnarray}
\label{Ser000}
\lefteqn{E_{000}(F) = -\frac{1}{2} - \frac{9}{4}\,F^2 -
\frac{3\,555}{64} \, F^4}
\nonumber\\[2ex]
& & \quad - \frac{2\,512\,779}{512} \, F^6 -
\frac{13\,012\,777\,803}{16~384} \, F^8 + \dots
\end{eqnarray}
The perturbation series for the state $n_1 = 3$, $n_2 = 0$, $m = 1$ is
alternating, but has a subleading nonalternating component [see
Eq.~(\ref{LargeN})]. The first perturbative terms read
\begin{eqnarray}
\label{Ser301}
\lefteqn{E_{301}(F) = -\frac{1}{50} + \frac{45}{2}\,F -
\frac{31875}{2} \, F^2}
\nonumber\\[2ex]
& & \quad + \frac{54\,140\,625}{4} \, F^3 -
\frac{715\,751\,953\,125}{16} \, F^4 + \dots
\end{eqnarray}
Note that for $F = 0$, the unperturbed nonrelativistic energy is
recovered, which is $-1/(2\,n^2)$ in atomic units.  In contrast to the
real perturbative coefficients, the energy pseudoeigenvalue (resonance)
$E(n_1,n_2,m,F)$ has a real and an imaginary component,
\begin{equation}
\label{ComplexEnergy}
E(n_1,n_2,m,F) = {\rm Re}\,E_{n_1 n_2 m}(F) -
\frac{i}{2} \, \Gamma_{n_1 n_2 m}(F)\,,
\end{equation}
where $\Gamma_{n_1 n_2 m}(F)$ is the autoionization width.

Using a computer algebra system~\cite{Wo1988,Disclaimer}, the first
50 nonvanishing perturbative
coefficients are evaluated for the ground state,
and for the state with parabolic quantum numbers
$n_1 = 3$, $n_2 = 0$, $m = 1$, we evaluate the first
70 nonvanishing perturbative
coefficients.
The apparent convergence of the
transforms defined in Eqs.~(\ref{ContourIntegral})
and~(\ref{AccelTrans}) in higher order is examined. In the case
of the Borel--Pad\'{e} transforms defined in Eq.~(\ref{ContourIntegral}),
use is made of the replacements
in Eqs.~(\ref{replacement000}) and~(\ref{replacement301})
[``leading renormalon poles are being put in by hand''].
This procedure leads to the numerical results listed in
Tables~\ref{table1} and~\ref{table2}.
The numerical error of our results is estimated on the basis
of the highest and lowest value of the four
highest-order transforms.

An important result of the comparison of the methods introduced in
Secs.~\ref{sec_bp} and~\ref{sec_cm} is the following: Both methods
appear to accomplish a resummation of the perturbation series to the
physically correct result.  Method I (Borel$+$Pad\'{e} with
leading renormalon poles, see
Sec.~\ref{sec_bp}) and method II (Borel$+$Pad\'{e}-improved
conformal mapping, see Sec.~\ref{sec_cm}) appear to lead to results
of comparable accuracy.

To date, a rigorous theory of the performance of the resummation
methods for divergent series of the type discussed in this work (with
alternating and nonalternating components) does not exist.
The {\em logarithmic} singularities introduced by
the branch points of higher-order polylogarithms
[see the index $k$ in Eq.~(\ref{polylog})] are difficult to
approximate with the rational functions employed in the construction of
Pad\'{e} approximants. A solution to the problem of approximating
the logarithmic singularities, based on finite number of perturbative
coefficients, would probably lead to further optimizimation of the rate of
convergence of the transformed series. Within
the current scheme of evaluation, the problematic logarithmic
singularities may be responsable, at least in part,
for certain numerical instabilities
at higher transformation order, e.g.~in the result for
${\mathcal T}''\!E_{70}(F = 2.1393\times10^{-4})$ in
Eq.~(\ref{method1}) below.

For the atomic state with quantum numbers $n_1 = 3$, $n_2=0$ and $m=1$, the
evaluation of the transforms ${\mathcal T}\!E_M(F)$ defined in
Eq.~(\ref{ContourIntegral}) (method I) and of the transforms ${\mathcal
  T}''\!E_M(F)$ defined in Eq.~(\ref{AccelTrans}) (method II) in
transformation order $M=67, 68, 69, 70$ for a field strength of
$F=2.1393\times10^{-4}$.  Method I leads to the following results,
\begin{eqnarray}
\label{method1}
& & {\mathcal T}\!E_{67}(F = 2.1393\times10^{-4}) =
  -0.015\,860\,468\,199~2\nonumber\\[1ex]
& & \qquad\qquad\qquad\qquad\qquad
- {\rm i} \; 0.529~048 \times 10^{-6} \,, \nonumber\\[2ex]
& & {\mathcal T}\!E_{68}(F = 2.1393\times10^{-4}) =
  -0.015\,860\,468\,200~9\nonumber\\[1ex]
& & \qquad\qquad\qquad\qquad\qquad
- {\rm i} \; 0.529~047 \times 10^{-6} \,, \nonumber\\[2ex]
& & {\mathcal T}\!E_{69}(F = 2.1393\times10^{-4}) =
  -0.015\,860\,468\,198~9\nonumber\\[1ex]
& & \qquad\qquad\qquad\qquad\qquad
- {\rm i} \; 0.529~048 \times 10^{-6}
\quad \mbox{and} \nonumber\\[2ex]
& & {\mathcal T}\!E_{70}(F = 2.1393\times10^{-4}) =
  -0.015\,860\,468\,194~5\nonumber\\[1ex]
& & \qquad\qquad\qquad\qquad\qquad
- {\rm i} \; 0.529~015 \times 10^{-6} \,.
\end{eqnarray}
%                

%
% table1
%
\begin{table}[htb]
\caption{Real and imaginary 
part of the energy pseudoeigenvalue $E_{000}(F)$
for the ground state of atomic hydrogen
(parabolic quantum numbers
$n_1 = 0, n_2 = 0, m = 0$).}
\label{table1}
\squeezetable
\begin{tabular}{lll}
\multicolumn{1}{c}{\rule[-3mm]{0mm}{8mm} $F$ (a.u.)} &
\multicolumn{1}{c}{\rule[-3mm]{0mm}{8mm} ${\rm Re}\,E_{000}(F)$} &
\multicolumn{1}{c}{\rule[-3mm]{0mm}{8mm} $\Gamma_{000}(F)$} \\
\hline
\rule[-3mm]{0mm}{8mm}
%
% F = 0.04
%
 $0.04$ &
 $-0.503~771~591~013~654~2(5)$ &
 $3.892~699~990(1) \times 10^{-6}$ \\
%
% F = 0.06
%
\rule[-3mm]{0mm}{8mm}
 $0.06$ &
 $-0.509~203~451~088(2)$ &
 $5.150~775~0(5) \times 10^{-4}$ \\
%
% F = 0.08
%
\rule[-3mm]{0mm}{8mm}
 $0.08$ &
 $-0.517~560~50(5)$ &
 $4.539~63(5) \times 10^{-3}$ \\
%
% F = 0.10
%
\rule[-3mm]{0mm}{8mm}
 $0.10$ &
 $-0.527~419~3(5)$ &
 $1.453~8(5) \times 10^{-2}$ \\
%
% F = 0.12
%
\rule[-3mm]{0mm}{8mm}
 $0.12$ &
 $-0.537~334(5)$ &
 $2.992~7(5) \times 10^{-2}$ \\
%
% F = 0.16
%
\rule[-3mm]{0mm}{8mm}
 $0.16$ &
 $-0.555~24(5)$ &
 $7.131(5) \times 10^{-2}$ \\
%
% F = 0.20
%
\rule[-3mm]{0mm}{8mm}
 $0.20$ &
 $-0.570~3(5)$ &
 $1.212(5) \times 10^{-1}$ \\
%
% F = 0.24
%
\rule[-3mm]{0mm}{8mm}
 $0.24$ &
 $-0.582~6(1)$ &
 $1.767(5) \times 10^{-1}$ \\
%
% F = 0.28
%
\rule[-3mm]{0mm}{8mm}
 $0.28$ &
 $-0.591~7(5)$ &
 $2.32(3) \times 10^{-1}$ \\
%
% F = 0.32
%
\rule[-3mm]{0mm}{8mm}
 $0.32$ &
 $-0.600(5)$ &
 $2.92(3) \times 10^{-1}$ \\
%
% F = 0.36
%
\rule[-3mm]{0mm}{8mm}
 $0.36$ &
 $-0.604(5)$ &
 $3.46(3) \times 10^{-1}$ \\
%
% F = 0.40
%
\rule[-3mm]{0mm}{8mm}
 $0.40$ &
 $-0.608(5)$ &
 $4.00(5) \times 10^{-1}$ \\
\end{tabular}
\end{table}

Method II yields the following data,
\begin{eqnarray}
\label{method2}
& & {\mathcal T}''\!E_{67}(F = 2.1393\times10^{-4}) =
  -0.015\,860\,468\,200~4\nonumber\\[1ex]
& & \qquad\qquad\qquad\qquad\qquad
- {\rm i} \; 0.529~047 \times 10^{-6} \,, \nonumber\\[2ex]
& & {\mathcal T}''\!E_{68}(F = 2.1393\times10^{-4}) =
  -0.015\,860\,468\,200~3\nonumber\\[1ex]
& & \qquad\qquad\qquad\qquad\qquad
- {\rm i} \; 0.529~047 \times 10^{-6} \,, \nonumber\\[2ex]
& & {\mathcal T}''\!E_{69}(F = 2.1393\times10^{-4}) =
  -0.015\,860\,468\,200~4\nonumber\\[1ex]
& & \qquad\qquad\qquad\qquad\qquad
- {\rm i} \; 0.529~047 \times 10^{-6} \quad \mbox{and} \nonumber\\[2ex]
& & {\mathcal T}''\!E_{70}(F = 2.1393\times10^{-4}) =
  -0.015\,860\,468\,203~3\nonumber\\[1ex]
& & \qquad\qquad\qquad\qquad\qquad
- {\rm i} \; 0.529~046 \times 10^{-6} \,. 
\end{eqnarray}
Numerical results obtained by resummation are presented
in Tables~\ref{table1} and~\ref{table2} for a variety of 
field strengths and for the two atomic states under investigation here. 
Results are compared to 
the numerical calculation~\cite{DaKo1976}.
which yields very accurate data for all experimentally accessible
electric field strengths to date. In addition, it should
be noticed that the inaccuracies at excessively large field
of~\cite{DaKo1976} have been pointed out by the same
authors in~\cite{Ko1987}.
However, not all atomic states considered in~\cite{DaKo1976}
were treated in the later investigation~\cite{Ko1987}.
Our data for the ground state indicated in Table~\ref{table1}
are consistent with the numerical results obtained in~\cite{Ko1987}.
However, it should be noted that the later work~\cite{Ko1987}
leaves out the excited state with quantum numbers
$n_1 = 3$, $n_2 = 0$ and $m = 1$ for which results
are given here here in Table~\ref{table2}.
To the best of our knowledge, the numerical discrepancy
with~\cite{DaKo1976} for the
excited state with quantum numbers $n_1 = 3$, $n_2 = 0$ and $m = 1$
has not been recorded in the literature. We do not claim here
that it would have been impossible to discern this discrepancy
with the other methods which have been devised for the
theoretical LoSurdo--Stark problem. Notably, it appears likely
that the approach from~\cite{Ko1987}
or the method presented in~\cite{BeGr1980} could easily be
generalized to the particular excited state considered here,
and that such a generalization would lead to very accurate
results. We merely include Table~\ref{table2} here in order to
illustrate the utility of the rather unconventional
resummation method for the regime
of large coupling, where even rather sophisticated numerical
methods, which avoid the intricacies of a small-field perturbative
expansion, have been shown to be problematic~\cite{DaKo1976,Ko1987}.
We confirm that the numerical data given 
in~\cite{DaKo1976} are accurate up to
a field strength of $F \approx 0.1$ for the ground state
and up to $F \approx 3 \times 10^{-4}$ for the excited ($n=5$)-state
with $n_1 = 3$, $n_2 = 0$ and $m = 1$.

\widetext
%
% table2
%
\begin{table}[tbh]
\caption{Real part and imaginary part
of the energy pseudoeigenvalue $E_{301}(F)$
for the excited state with parabolic quantum numbers
$n_1 = 3, n_2 = 0, m = 1$. The field strength $F$ is given
in atomic units. The data are compared
to Ref.~\protect\cite{DaKo1976}. Discrepancies are observed at 
large electric field strength.}
\label{table2}
\squeezetable
\begin{tabular}{l@{\hspace*{1.5cm}}l@{\hspace*{0.5cm}}%
l@{\hspace*{1.5cm}}l@{\hspace*{0.5cm}}l}
\multicolumn{1}{c}{} &
\multicolumn{2}{c}{\rule[-3mm]{0mm}{8mm} Real part of the resonance
${\rm Re}\,E_{301}(F)$} &
\multicolumn{2}{c}{\rule[-3mm]{0mm}{8mm}
Autoionization decay width $\Gamma_{301}(F)$} \\
\multicolumn{1}{l}{\rule[-3mm]{0mm}{8mm} $F$ (a.u.)} &
\multicolumn{1}{c}{Ref.~\protect\cite{DaKo1976}} &
\multicolumn{1}{c}{Our results} &
\multicolumn{1}{c}{Ref.~\protect\cite{DaKo1976}} &
\multicolumn{1}{c}{Our results} \\
\hline
%
% F = 1.5560 {-4}
%
\rule[-3mm]{0mm}{8mm}
 $1.5560 \times 10^{-4}$ &
 $-0.016~855~237~2$ &
 $-0.016~855~237~140~761~7(5)$ &
 $0.42 \times 10^{-9}$ &
 $0.421~683(5)\times 10^{-9}$ \\
%
% F = 1.9448 {-4}
%
\rule[-3mm]{0mm}{8mm}
 $1.9448 \times 10^{-4}$ &
 $-0.016~179~388~5$ &
 $-0.016~179~388~257~0(5)$ &
 $0.143~8 \times 10^{-6}$ &
 $0.143~773(5) \times 10^{-6}$ \\
%
% F = 2.1393 {-4}
%
\rule[-3mm]{0mm}{8mm}
 $2.1393 \times 10^{-4}$ &
 $-0.015~860~468$ &
 $-0.015~860~468~20(1)$ &
 $0.105~7 \times 10^{-5}$ &
 $0.105~09(5) \times 10^{-5}$ \\
%
% F = 2.5282 {-4}
% 
\rule[-3mm]{0mm}{8mm}
 $2.5282 \times 10^{-4}$ &
 $-0.015~269~204$ &
 $-0.015~269~293(1)$ &
 $0.175~60 \times 10^{-4}$ &
 $0.176~39(5) \times 10^{-4}$ \\
%
% F = 2.9172 {-4}
%
\rule[-3mm]{0mm}{8mm}
 $2.9172 \times 10^{-4}$ &
 $-0.014~740~243$ &
 $-0.014~742~60(3)$ &
 $0.976~51 \times 10^{-4}$ &
 $0.999~96(9)\times 10^{-4}$ \\
%
% F = 3.3061 {-4}
%
\rule[-3mm]{0mm}{8mm}
 $3.3061 \times 10^{-4}$ &
 $-0.014~242~49$ &
 $-0.014~260~2(3)$ &
 $0.278~53 \times 10^{-3}$ &
 $0.295~4(2) \times 10^{-3}$ \\
\end{tabular}
\end{table}

\newpage

\narrowtext

\begin{center}
\begin{minipage}{8.0cm}
\begin{table}[tbh]
\caption{\label{sailer} Resummation of the asymptotic series for the
generating functional of a zero-dimensional theory with degenerate
minima given in Eqs.~(\ref{powerg}) and~(\ref{divexp}).
We have $g = 0.01$.
Results in the third column are obtained by the method indicated in
Eq.~(\ref{stenmark}) along the integration contour $C_{0}$
(see~\protect\cite{Je2000prd}).  The
partial sums in the second column are obtained from the asymptotic
series~(\ref{powerg}).}
\begin{center}
\begin{minipage}{8.0cm}
\begin{tabular}{cll}
\multicolumn{1}{c}{\rule[-3mm]{0mm}{8mm}{$M$}} &
\multicolumn{1}{c}{\rule[-3mm]{0mm}{8mm}{partial sum}} &
\multicolumn{1}{c}{\rule[-3mm]{0mm}{8mm}{${\mathcal T}Z_M(g = 0.01)$}} \\
\hline
2  &
$1.081~000$ &
$1.102~326$ \\
3  &
$1.094~860$ &
$1.096~141$ \\
4  &
$1.108~373$ &
$1.089~875$ \\
5  &
$1.125~832$ &
$1.090~695$ \\
6  &
$1.153~942$ &
$1.092~000$ \\
7  &
$1.208~154$ &
$1.091~596$ \\
8  &
$1.329~994$ &
$1.091~389$ \\
9  &
$1.642~718$ &
$1.091~553$ \\
10 &
$2.545~239$ &
$1.091~545$ \\
11 &
$5.438~230$ &
$1.091~503$ \\
12 &
$1.5 \times 10^1$ &
$1.091~525$ \\
13 &
$5.5 \times 10^1$ &
$1.091~527$ \\
14 &
$2.2 \times 10^2$ &
$1.091~519$ \\
15 &
$9.5 \times 10^2$ &
$1.091~523$ \\
16 &
$4.5 \times 10^3$ &
$1.091~523$ \\
17 &
$2.2 \times 10^4$ &
$1.091~521$ \\
18 &
$1.2 \times 10^5$ &
$1.091~522$ \\
19 &
$6.9 \times 10^5$ &
$1.091~522$ \\
20 &
$4.1 \times 10^6$ &
$1.091~522$ \\
\hline
\multicolumn{1}{l}{exact} &
$1.091~522$ &
$1.091~522$ \\
\end{tabular}
\end{minipage}
\end{center}
\end{table}
\end{minipage}
\end{center}

\typeout{Section 6}
%
% MODEL EXAMPLE FOR DEGENERATE MINIMA 
%
\section{MODEL EXAMPLE FOR DEGENERATE MINIMA} 
\label{degen}

We consider the generating functional in a zero-dimensional
theory (in this case, the 
usual path integral reduces to an ordinary integral).
First, we briefly consider the $\Phi^4$-theory in zero
dimensions [see Eq.~(9-177) ff.~in~\cite{ItZu1980}];
the generating functional reads 
\begin{equation}
Z(\Phi) = \int\limits_{-\infty}^{\infty}
\frac{{\rm d} \Phi}{\sqrt{2 \pi}} \, \exp\left[ - \frac{1}{2} \, \Phi^2 \, 
- g \, \Phi^4 \right] \,.
\end{equation}
The strictly alternating
divergent asymptotic expansion in powers of $g$ for
$g \to 0$ reads,
\begin{equation}
Z(\Phi) \sim \sum_{N=0}^{\infty} 
\frac{4^{N} \, \Gamma(2 N + 1/2)}
{\sqrt{\pi} \, \Gamma(N + 1)} \, (-g)^N\,.
\end{equation}
On using the known asymptotics valid 
for $N \to \infty$, which is this case 
yield the ``large-order'' asymptotics
of the perturbative coefficients,
\begin{equation}
\label{asympgamma}
\frac{\Gamma(2 N + 1/2)}{\Gamma(N + 1)} \sim
\frac{4^N}{\sqrt{2 \pi}} \, \Gamma(N) \, \left[ 1 + 
{\mathcal O}\left(\frac{1}{N}\right) \right] \,
\end{equation}
it is easy to explicitly establish the factorial divergence
of the series~(see also p.~888 of~\cite{ZJ1996}).
The generating functional in zero dimensions has been
proposed as a paradigmatic example for the divergence of perturbation
theory in higher order. It can be resummed easily to the nonperturbative
result; in particular it is manifestly Borel summable, and no
singularities are present on the positive real axis.

Complications are introduced
by degenerate minima. As a second example, we consider the 
modified generating functional [compare with
Eq.~(2.6) on p.~15 of~\cite{LGZJ1990} and with
Eq.~(40.1) on p.~854 of~\cite{ZJ1996}]:
\begin{eqnarray}
\label{zp}
Z'(\Phi) &=& \int\limits_{-\infty}^{\infty}
\frac{{\rm d} \Phi}{\sqrt{2 \pi}} \, \exp\left[ - \frac{1}{2} \, \Phi^2 \, 
(1 - \sqrt{g} \, \Phi)^2 \right] \nonumber\\[2ex]
&=& \int\limits_{-\infty}^{\infty}
\frac{{\rm d} \Phi}{\sqrt{2 \pi}} \, \exp\left[ - \frac{1}{2} \, \Phi^2 + 
\sqrt{g} \, \Phi^3  - \frac{1}{2} \, g \, \Phi^4 \right] \,.
\end{eqnarray}
The expansion of the exponential in powers of the coupling
$g$ leads to a divergent asymptotic series,
\begin{eqnarray}
\label{powerg}
Z'(\Phi) &=& \sum_{N=0}^{\infty} \frac{1}{N!} \,
\int\limits_{-\infty}^{\infty}
\frac{{\rm d} \Phi}{\sqrt{2 \pi}} \,\, {\rm e}^{- 1/2 \, \Phi^2} \,
\left(\sqrt{g} \, \Phi^3  - \frac{1}{2} \, g \, \Phi^4\right)^N
\nonumber\\[3ex]
&=& \sum_{N=0}^{\infty} \,\,
\int\limits_{-\infty}^{\infty}
\frac{{\rm d} \Phi}{\sqrt{2 \pi}} \,\, {\rm e}^{- 1/2 \, \Phi^2} 
\sum_{j=0}^{N} \frac{(-1)^j}{\Gamma(2 N - j + 1)} 
\nonumber\\
& & \quad \times {2 N - j \choose j} \,
\left( \sqrt{g} \, \phi^3 \right)^{2 (N - j)} \,
\left( \frac{g \, \phi^4}{2} \right)^j 
\nonumber\\[3ex]
&=& \sum_{N=0}^{\infty} \, 2 \sqrt{\pi} \,
\frac{(-1)^N \, C_{2 N}^{N + 1/2}(1)}{\Gamma(N - 1/2)} \, g^N
\nonumber\\[3ex]
&=& \sum_{N=0}^{\infty} 
\frac{8^N \, \Gamma(2 N + 1/2)}{\sqrt{\pi} \, \Gamma(N + 1)} \, g^N\,,
\end{eqnarray}
where $C_M^N(x)$ denotes a Gegenbauer (ultraspherical) polynomial. 
Note that terms of half-integer
power of $g$ entail an odd power of the 
field and vanish after integration. The first few terms
of the asymptotic expansion read,
\begin{eqnarray}
\label{divexp}
\lefteqn{Z'(\Phi) = 1 + 6 \, g +
210 \, g^2 + 13860 \, g^3 }\nonumber\\[2ex]
& & \quad + 1351350 \, g^4 +
174594420 \, g^5 \nonumber\\[2ex]       
& & \quad + 28109701620 \, g^6    
+ 5421156741000 \, g^7 \nonumber\\[2ex]
& & \quad + 1218404977539750 \, g^8 + 
{\mathcal O}(g^9) \,. 
\end{eqnarray}
For the perturbative coefficients
\begin{equation}
C_N = \frac{8^N \, \Gamma(2 N + 1/2)}{\sqrt{\pi} \, \Gamma(N + 1)} \,,
\end{equation}
we establish the following asymptotics,
\begin{equation} 
\label{asympg}
C_N \sim \frac{1}{\pi \sqrt{2}} \, N^{-1} \, 32^N \, \Gamma(N+1) \,.
\end{equation}
Due to the nonalternating character of the expansion (\ref{powerg}),
it is not Borel summable in the ordinary sense. 
Rather, it is Borel summable in the 
distributional sense~\cite{CaGrMa1986,CaGrMa1993}.
Here, we present numerical evidence supporting the 
summability of the divergent expansion~(\ref{divexp}) 
based on a finite number of perturbative coefficients. 
The final integration is carried out along the 
contour $C_0$ introduced in~\cite{Je2000prd}
[see also Eq.~(\ref{stenmark}) below].
The same contour has also been used for the resummation of
divergent perturbation series describing renormalization
group (anomalous dimension) $\gamma$ functions~\cite{JeSo2000hep}.
As explained in~\cite{Je2000prd}, the integration along
$C_0$, which is based on the mean value of the results
obtained above and below the real axis,
leads to a {\em real} final result if all perturbative
coefficients are real.

In particular, the resummation of the divergent expansion 
(\ref{divexp}) is accomplished as follows.
We first define the Borel transform of the 
generating functional by [see Eq.~(4) in~\cite{JeWeSo2000}
and the discussion after Eq.~(\ref{BorelTrans})]
\begin{eqnarray}
\label{ZBorel}
Z'_{\rm B}(z) & \equiv &  
{\mathcal B}^{(1,1)}\left[Z'; \, z\right] 
\nonumber\\[1ex]
& = & \sum_{N=0}^{\infty}
\frac{C_N}{\Gamma(N + 1)} \, z^N\,.
\end{eqnarray}
Pad\'{e} approximants to this Borel transform are evaluated, 
\begin{equation}
{\mathcal P'}_M(z) = \bigg[ [\mkern - 2.5 mu [M/2] \mkern - 2.5 mu ] \bigg/
[\mkern - 2.5 mu [(M+1)/2] \mkern - 2.5 mu ]
\bigg]_{Z'_{\rm B}}\!\!\!\left(z\right)\,,
\end{equation}
where $[\mkern - 2.5 mu [x] \mkern - 2.5 mu ]$ denotes the largest
positive integer smaller than $x$.  We then evaluate the (modified)
Borel integral along the integration contour $C_{0}$ 
in troduced in~\cite{Je2000prd}; specifically we define
the transform ${\mathcal T}\!Z_M(g)$
\begin{equation}
\label{stenmark}
{\mathcal T}\!Z_M(g) = \int_{C_{0}} {\rm d}t \, 
\exp(-t)\,{\mathcal P'}_M(g\,t) \,.
\end{equation}
In this case, poles above and below the real axis must be considered,
and the final result involves no imaginary part. 
The particular case of $g = 0.01$ is considered. Values for
the partial sums of the perturbation series (\ref{divexp}) and the
transforms defined in Eq.~(\ref{stenmark}) are 
shown in Tab.~\ref{sailer}. The transforms exhibit apparent
convergence to 6 decimal places in 20th order, whereas the 
partial sums of the perturbation series diverge.
Between the second and forth term of the perturbation series,
(the forth term constitutes the minimal term), the partial sums
provide approximations to the exact result. 
It might seem surprising that the minimal term in the perturbative
expansion is reached already in forth order, although the 
coupling assumes the small value $g = 0.01$. This behavior
immediately follows from the large geometric factor 
in Eq.~(\ref{asympg}) which leads to a 
``resultative coupling strength parameter''
of $g_{\rm res} = 0.32$. ``Nonperturbative effects'' of the order
of $\exp(-1/g_{\rm res})$ provide a fundamental limit to the accuracy
obtainable by optimal truncation of the perturbation series;
this is consistent with the numerical data in Table~\ref{sailer}.

We have also investigated the 
resummation of the divergent series (\ref{divexp}) via a
combination of a conformal mapping and Pad\'{e} aproximants
in the conformal variable. The situation is analogous to the
LoSurdo--Stark effect: Results are consistent than those
presented in Table~\ref{sailer} obtained by the
``pure'' Borel--Pad\'{e} and in this case slightly
more accurate.
The radius of convergence of the Borel transform $Z'_{\rm B}(z)$
defined in Eq.~(\ref{ZBorel}) is $s = 1/32$
[cf.~Eq.~(\ref{defS}) for the LoSurdo--Stark effect], and the 
appropriate conformal mapping in this case reads
\begin{equation}
\label{conformal3}
w = \frac{4\,y}{(1 + y)^2}
\end{equation}
[cf.~Eq.~(\ref{conformal1})]. The inverse reads
\[
y(w) = \frac{1 - \sqrt{1 - w}}{1 + \sqrt{1 - w}}
\]
[cf.~Eq.~(\ref{conformal2})]. The conformal mapping (\ref{conformal3})
maps the complex $w$-plane with a cut along $(1,\infty)$
unto the unit circle in the complex $y$-plane.
While the zero-dimensional model example given in Eq.~(\ref{zp}) does
not exhibit all problematic features of
degenerate anharmonic double-well oscillators,
certain analogies can be established; these comprise
in particular the need to evaluate the mean value of Borel
transforms above and below the real axis
(see Appendix~\ref{appb}).

\typeout{Section 7}
%
% Conclusion
%
\section{CONCLUSION}
\label{sec_co}

We discuss the resummation of the divergent perturbation
series of the LoSurdo--Stark effect,
and of a divergent model series describing a zero-dimensional theory
with degenerate minima, using two methods.
Method I, which uses a variant of the 
contour-improved Borel--Pad\'{e} technique introduced
in~\cite{FrGrSi1985}, is described in Sec.~\ref{sec_bp}. 
The integration contour is modified
so that the additional terms which have to be added to the
principal value of the Laplace--Borel integral are clearly
identified [see also the discussion in~\cite{Je2000prd} and 
Eq.~(\ref{ContourIntegral2})]. Use is made of the
leading infrared renormalon pole. 
Method II, which comprises an analytic continuation by conformal
mapping with additional improvement by Pad\'{e} approximants
in the conformal variable [see Eq.~(\ref{ConformalPade})],
is discussed in Sec.~\ref{sec_cm}.
This method is a variant of the method introduced 
in~\cite{CaFi1999,CaFi2000} which has been 
shown to accelerate convergence
of perturbative quantum chromodynamics (by optimal conformal
mapping of the Borel plane). We find that 
{\em both} methods accomplish a resummation of the divergent
perturbation series (\ref{PertSer}) for the LoSurdo--Stark effect,
and the decay width of the quasistationary states is
obtained~(see Sec.~\ref{sec_nc}. 
Numerical results are given in Tables~\ref{table1} and~\ref{table2}).
A main result of the current investigation
is the demonstration of the analogous mathematical structure
(doubly-cut Borel plane)
of the perturbative expansion for the LoSurdo--Stark effect
and perturbative expansions in quantum chromodynamics
(renormalon theory~\cite{Be1999}).
The series investigated here exhibit a nontrivial singularity
structure in the Borel plane. In particular, we encounter
poles and branch cuts on the positive real axis.

In quantum electrodynamics, we encounter nonperturbative
effects in the electron-positron pair-production amplitude
in a background electric 
field~\cite{HeEu1936,Sc1951,ItZu1980,GrRe1992}. 
The vacuum-to-vacuum amplitude
acquires an imaginary part, whose magnitude is related to 
the production rate per space-time volume of fermion-antifermion
pairs. This nonperturbative, imaginary contribution can be
inferred from the perturbative expansion of the effective
action by contour-improved resummation
(see~\cite{Je2000prd} and the discussion in Appendix~\ref{appa}).
Nonperturbative effects typically involve
a nonanalytic factor of $\exp(-1/g)$ where $g$ is an 
appropriate coupling parameter for the physical system
under investigation (in the case of the LoSurdo--Stark
effect, the coupling parameter is the electric field strength $F$). 
The existence of nonperturbative 
contributions is intimately linked
with the failure of the Carleman criterion for a particular
perturbation series
(see for example~\cite{GrGrSi1970}, Theorems XII.17 and XII.18 and the
definition on p.~43 in~\cite{ReSi1978}, 
p.~410 in~\cite{BeOr1978}, or the elucidating discussion
in Ref.~\cite{Fi1997}). 
The Carleman criterion determines,
roughly speaking, if nonanalytic contributions exist for a given
effect which is described by a specified perturbation series.

The current investigation illustrates the 
utility of resummation methods in those cases where
perturbation theory breaks down at large coupling. 
As explained in Sec.~\ref{sec_nc},
even in situations where the 
perturbation series diverges strongly, it can still be used
to obtain meaningful physical results if it is
combined with a suitable resummation method.
In a relatively weak field, it is possible to obtain more 
accurate numerical results by resummation
than by optimal truncation of the
perturbation series~(see also~\cite{Je2000prd}). In a strong field,
it is possible to obtain physically correct
results by resummation even though the perturbation series 
diverges strongly (see the discussion in Sec.~\ref{sec_nc}
and the data in Tables~\ref{table1},~\ref{table2} and~\ref{sailer}).
By resummation, the perturbation series which is 
inherently a weak-coupling expansion can be given a 
physical interpretation even
in situations where the coupling is large.
Returning to the analogy to quantum field theory,
one might be tempted to suggest that physically complete
results are obtained after regularization, renormalization
{\em and} resummation.

\section*{ACKNOWLEDGMENTS}

It is a pleasure to thank Professor G.~Soff and
Professor J.~Zinn--Justin for many insightful
discussions, and G.~Soff for continuous encouragement.
Helpful discussions with Professor P.~J.~Mohr, E.~Caliceti,
E.~J.~Weniger, J.~Sims,
I.~N\'{a}ndori and S.~Roether are also gratefully
acknowledged. The author would also like to acknowledge  
support from the Deutscher Akademischer Austauschdienst (DAAD)
during a post-doctoral fellowship, and he would like
to thank the Laboratoire Kastler--Brossel, where major parts
of this work were performed, and
the National Institute for Standards and Technology for kind hospitality.

\appendix

%
% Divergent Perturbation Series in Quantum Field Theory
%

\section{DIVERGENT PERTURBATION\\ SERIES IN QUANTUM FIELD THEORY}
\label{appa}

We briefly indicate aspects of certain divergent perturbation
series in quantum field theory, in particular the quantum
electrodynamic (QED) effective action and the associated
pair--production amplitude for electron--positron pairs. We use
natural units in which the reduced Planck's constant, the permittivity
of the vacuum and the speed of light (in field-free vacuum) assume the
value of unity ($\hbar = \epsilon_0 = c = 1$).  The one-loop QED
effective action for an arbitrary electric and magnetic field per
space-time volume reads [this result can be found e.g.~in Eq.~(4-123)
of~\cite{ItZu1980}, upon inclusion of an additional counterterm; 
see also~\cite{Sc1951,DiGi2000}]
\begin{eqnarray}
\label{Leff}
\lefteqn{S = \lim_{\epsilon,\eta \to 0^{+}}
- \frac{1}{8 \pi^2} \int\limits_0^{{\rm i}\infty + \eta}
\frac{{\rm d} s}{s^3} \,
{\rm e}^{- (m_{\rm e}^2 - {\rm i} \epsilon)\, s}}
\nonumber\\
& & \qquad \times \,
\biggl[ (e s)^2 \, a b \, \coth(e a s) \, \cot(e b s) \nonumber\\
& & \qquad \qquad - \frac{1}{3} \, (e s)^2 \, (a^2 - b^2) - 1 \biggr]\,.
\end{eqnarray}
The specification of the 
infinitesimal quantity $\eta$ is necessary, strictly
speaking, in order to avoid the singularities of the 
$\coth$ function along the imaginary axis.
By $a$ and $b$ we denote the {\em secular invariants},
\begin{eqnarray}
a &=& \sqrt{\sqrt{{\mathcal F}^2 + 
{\mathcal G}^2} + {\mathcal F}^2}\,,\nonumber\\
b &=& \sqrt{\sqrt{{\mathcal F}^2 + 
{\mathcal G}^2} - {\mathcal F}^2}\,,\nonumber\\
{\mathcal F} &=& \frac{1}{4} \, F^{\mu\nu} \, F_{\mu\nu} =
\frac{1}{2} \, (\bbox{B}^2 - \bbox{E}^2)\,, \nonumber\\
{\mathcal G} &=& \frac{1}{4} \, F^{\mu\nu} \, (* F)_{\mu\nu} =
- \bbox{E} \cdot \bbox{B}\,. \nonumber
\end{eqnarray}
If the relativistic invariant ${\mathcal G}$ is positive,
then it is possible to transform to a Lorentz frame in which $\bbox{E}$ and
$\bbox{B}$ are {\em antiparallel}. In the case ${\mathcal G} < 0$,
it is possible to choose a Lorentz frame in which  $\bbox{E}$ and
$\bbox{B}$ are {\em parallel}. Irrespective of the sign
of ${\mathcal G}$ we have in the specified frame
\[
a = |\bbox{B}| \quad \mbox{and} \quad b = |\bbox{E}|\,.
\]
Then, in the special Lorentz frame, the effective action reads
[we keep all infinitesimal contributions]
\begin{eqnarray}
\label{Leff2}
\lefteqn{S = \lim_{\epsilon,\eta \to 0^{+}}
- \frac{1}{8 \pi^2} \int\limits_0^{{\rm i}\infty + \eta}
\frac{{\rm d} s}{s^3} \,
{\rm e}^{- (m_{\rm e}^2 - {\rm i} \epsilon)\, s}} \,
\nonumber\\ 
& & \qquad \times \,
\biggl[ (e s)^2 \, |\bbox{B}| \, |\bbox{E}| \,
\coth(e |\bbox{B}| s) \, \cot(e |\bbox{E}| s) \nonumber\\
& & \qquad \qquad 
- \frac{1}{3} \, (e s)^2 \, (\bbox{B}^2 - \bbox{E}^2) - 1 \biggr]\,.
\end{eqnarray}
The particular cases of a pure magnetic and a pure electric field 
follow from the above integral representation by
considering appropriate limits ($|\bbox{E}| \to 0$ and
$|\bbox{B}| \to 0$, respectively).
These particular cases
are of some interest, because they can be used as model series for
divergent alternating and nonalternating asymptotic perturbation
series~\cite{DuHa1999,JeBeWeSo2000,Je2000prd,JeWeSo2000}.  In the case of
a pure magnetic field $B = |\bbox{B}|$, the result reads [see
e.g.~Eq.~(5) in~\cite{JeBeWeSo2000}]
\begin{equation}
\label{SRB}
S_B = - \frac{e^2 B^2}{8 \pi^2}
\int\limits_0^\infty \frac{{\rm d}s}{s^2} \,
\left\{\coth s - \frac{1}{s} - \frac{s}{3} \right\}
\exp\!\left(-\frac{m_{\rm e}^2}{e\,B} s\right)\,,
\end{equation}
where $m_{\rm e}$ is the mass of the electron. This result can be
expressed as a divergent asymptotic perturbation series in the
coupling parameter $g_B = e^2 B^2 / m_{\rm e}^4$. For the pure
electric field, the result reads
\begin{eqnarray}
\label{SRE}
S_E &=& \frac{e^2 E^2}{8 \pi^2}
\int\limits_0^\infty \frac{{\rm d}s}{s^2} \,
\left\{\coth s - \frac{1}{s} - \frac{s}{3} \right\} \nonumber\\
& & \qquad \qquad \times \exp\left[- {\rm i} \left(\frac{m_{\rm e}^2}{e\,E} -
{\rm i}\,\epsilon\right)\,s\right] \nonumber\\[2ex]
&=& - \frac{e^2 E^2}{8 \pi^2} \!
\int\limits_{0 + {\rm i}\,\epsilon}^{\infty + {\rm i}\,\epsilon} \!
\frac{{\rm d}s}{s^2} \!
\left\{\cot s \! - \frac{1}{s} \! + \! \frac{s}{3} \right\} 
\exp\left[-\frac{m_{\rm e}^2}{e\,E} s\right]
\end{eqnarray}
[Eq.~(7) of~\cite{Je2000prd} and the expression before Eq.~(10)
of~\cite{JeBeWeSo2000} contain typographical errors]. We take the
opportunity to supplement the proportionality factor for the
expression in Eq.~(7) of~\cite{Je2000prd} to yield the effective action
per space-time volume element; it reads $e^2 E^2/(8 \pi^2)$. As
evident from the Eq.~(\ref{SRE}), the integration of the
Borel--Pad\'{e} transform for the electric field case should be
carried out along the contour $C_{+1}$ shown here in
Fig.~\ref{figg1}. When this contour is used,
then a sign change results for the imaginary contributions in Table~1
of~\cite{Je2000prd} (the sign change of the imaginary part according to
the choice of the integration contour has been discussed at length
in~\cite{Je2000prd}).  The magnitude of the imaginary part yields the
pair-production amplitude. The contour $C_{+1}$ is used in the
current investigation (and in the context of the related brief
discussion in~\cite{Je2000prd}) for the calculation of nonperturbative
imaginary effects, i.e.~the autoionization decay width of atomic
states (LoSurdo--Stark effect).  

The divergent asymptotic perturbation series for the cases of the
magnetic and electric field, generated by expansion of the results in
Eqs.~(\ref{SRB}) and~(\ref{SRE}), can be found in Eqs.~(6) and (7)
of~\cite{JeBeWeSo2000} ($B$--field, alternating series, coupling
parameter $g_B = e^2 B^2 / m_{\rm e}^4$) and in Eqs.~(8) and (9)
of~\cite{Je2000prd} ($E$--field, nonalternating series, coupling
parameter $g_E = e^2 E^2 / m_{\rm e}^4$). The singularity structure of
the Borel transform of the series for the magnetic field case has been
determined in~\cite{JeWeSo2000} as a sequence
of singularities corresponding to 
alternating, factorially divergent components
(these correspond in their mathematical 
structure to the so-called ultraviolet renormalons in quantum
chromodynamics). The perturbation series
for the LoSurdo--Stark effect contains both nonalternating and
alternating components so that its resummation represents a
comparatively more interesting task. The same applies to the more
complex perturbation series calculated in~\cite{BrKr2000} for the
renormalization group $\gamma$ function, whose resummation -- at
strong coupling -- has been discussed in~\cite{JeSo2000hep,BrKr2000}
(in this case, there is no imaginary part involved). 
We are not aware of any {\em a priori} reasoning to determine the
absence or presence of imaginary contributions in a particular
physical problem (see also the discussion in~\cite{Fi1997polo}).

%
% BOREL SUMMABILITY IN PROBLEMATIC CASES}
% 
\section{BOREL SUMMABILITY\\ IN PROBLEMATIC CASES}
\label{appb}

Consider the one-dimensional double-well Hamiltonian
\begin{equation}
H(g) = p^2 + x^2 \, (1 - g\,x)^2\,.
\end{equation}
For $g = 0$, the Hamiltonian describes harmonic oscillations.
For $g > 0$, we have degenerate minima of the 
potential at $x = 0$ and at
$x = 1/g$, and to each eigenvalue of the unperturbed harmonic
oscillator, we have to associate two eigenvalues belonging
to opposite-parity wavefunctions with respect to $x = 1/(2 g)$.
The difference of the two eigenvalues is nonanalytic in $g$.
Two different approaches have been developed to circumvent this
problem and to allow for a treatment based on the resummation
of perturbation series.

The first approach~\cite{ZJ1996} is based essentially on 
generalized Bohr--Sommerfeld quantization formulas and leads
to an expansion of the ground-state energy eigenvalue in terms of
\begin{eqnarray}
\lefteqn{E(g) = \sum_{n=0}^{\infty} E^{(0)}_l \, g^l +
\sum_{n=1}^{\infty} \frac{1}{\sqrt{\pi \, g}} \,
{\rm e}^{-1/6g}} \nonumber\\
& & \quad \times \sum_{k=0}^{n-1} [\ln(-2/g)]^k \,
\sum_{l=0}^{\infty} E^{(n)}_{kl} \, g^l\,,
\end{eqnarray}
where the upper index of the $E$-coefficients labels the
multi-instanton contributions (the zero-instanton contribution
corresponds to the ``usual'' perturbative expansion).
The odd-instanton contributions lead to a separation of the 
ground state and the first excited state which have opposite
parity but the same naive perturbation series 
$\sum_{n=0}^{\infty} E^{(0)}_l \, g^l$.
The series $\sum_{n=0}^{\infty} E^{(0)}_l \, g^l$ is 
nonalternating and therefore not Borel-summable; however, possible
imaginary contributions must be suppressed for physical reasons because
the ground-state energy is real. The suppression can be enforced 
explicitly by defining the sum as the arithmetic
mean of the values obtained
above and below the real axis, or it can be motivated
by the following observation~\cite{ZJ1996}:
We define the sum of $\sum_{n=0}^{\infty} E^{(0)}_l \, g^l$
for negative $g$ and carry out an analytic continuation to positive
$g$; this leads to an imaginary part which in this case
cancels with the imaginary part generated by the $\ln(-2/g)$ coming
from the two-instanton contribution ($n=2$, $k=1$). Note that for 
the model example
discussed in Sec.~\ref{degen}, only one alternative is feasible
-- the explicit cancellation --, because no additional terms
are present which could lead to cancellations. 

The second approach~\cite{CaGrMa1988,CaGrMa1996}
involves contour integrations and makes use of
projection operators in order to ``select'' states
of specified parity
(this approach has
been shown to be
applicable as well to the problematic 
Herbst--Simon hamiltonian~\cite{HeSi1978plb} which
involves a vanishing perturbation series).
Specifically, we can write the perturbed energy eigenvalue as
\begin{equation}
E(g) = \frac{{\rm Re} F_1(g,g)}{{\rm Re} F_0(g,g)}
\end{equation}
where
\begin{equation}
F_j (g,\gamma) = \frac{1}{2\,\pi\,{\rm i}} \oint_\Gamma
{\rm d} z \, z^j \, \langle \psi^{\pm}(g) | \frac{1}{H(\gamma) - z} |
\psi^{\pm}(g) \rangle \,.
\end{equation}
where $\psi^{\pm}(g)$ are {\em test functions} with a definite parity
with respect to $1/(2 g)$. The closed contour $\Gamma$ has radius unity;
it is chosen to encircle one and only one 
shifted resonance of the perturbed oscillator,
while the test functions select the state with the 
desired parity. Specifically, we have 
$\psi^{\pm}(g) = P^\pm(g) \psi$ where $\psi$ is the eigenvector of the 
unperturbed hamiltonian, and the projection operators are
\begin{equation}
\left(P^\pm(g) \psi\right)(x) = \frac{1}{2} \, \left( \psi(x)  \pm
\psi(1/g - x) \right)\,.
\end{equation}
The functions $F_j(g,\gamma)$ may be expressed as asymptotic
series,
\begin{equation}
F_j (g,\gamma) = \sum_{N=0}^M \left[ a_{j,N} + {\rm i} \, b_{j,N}(g)
\right] \, \gamma^N + {\mathcal O}(\gamma^{M+1})\,.
\end{equation}
The authors of~\cite{CaGrMa1988,CaGrMa1996}
define $\Phi_j^{\rm R}(g,\gamma)$ to be the Borel sum of 
$\sum_{N=0}^\infty a_{j,N} \, \gamma^N$ for $g$,
$|\gamma|$ and $\arg \gamma$ small and positive,
and they establish a corresponding relation 
for $\Phi_j^{\rm I}(g,\gamma)$ and
$\sum_{N=0}^\infty b_{j,N} \, \gamma^N$. 
According to (unnumbered) equations on p.~626
of~\cite{CaGrMa1988},
the final result in this case is given in terms of the mean values
-- each obtained above and below the real axis -- of 
the two Borel sums $\Phi_j^{\rm R}$ and $\Phi_j^{\rm I}$, 
\begin{eqnarray}
{\rm Re}F_j(g, \gamma) & = & \frac{1}{2} \, 
\left[ \Phi_j^{\rm R}(g,\gamma) + 
\overline{\Phi_j^{\rm R}(g,{\overline \gamma})} \right]
\nonumber\\
& & \quad + \frac{\rm i}{2} \, \left[ \Phi_j^{\rm I}(g,\gamma) -
\overline{\Phi_j^{\rm I}(g,{\overline \gamma})} \right] \,,
\end{eqnarray}
where ${\overline z}$ denotes the complex conjugate of $z$.
The value ${\rm Re}F_j(g, g)$ is then determined by (unique)
analytic continuation $\gamma \to g$ from
${\rm Re}F_j(g, \gamma)$. 
In our simplified model example,
we have $b_{j,N}(g) = 0$ (all perturbative coefficients are real). 
The need to evaluate the arithmetic mean of Borel sums
above and below the real axis appears to arise naturally in the 
context of double-well problems.

\widetext


\begin{thebibliography}{10}
\bibitem{Disclaimer}
Certain commercial equipment, instruments, or materials are identified in this
  paper to foster understanding. Such identification does not imply
  recommendation or endorsement by the National Institute of Standards and
  Technology, nor does it imply that the materials or equipment identified are
  necessarily the best available for the purpose.
\end{thebibliography}
\end{document}